\documentclass[aps,prb,superscriptaddress]{revtex4-2}
\usepackage{graphicx} 
\usepackage{float}
\usepackage{booktabs}  
\usepackage{xcolor}
\usepackage{subcaption}
\usepackage{amsmath}

\usepackage{multirow}
\usepackage{rotating}
\usepackage[T1]{fontenc}
\usepackage{parskip}
\usepackage{array} 

\usepackage{adjustbox}

\begin{document}
\title{\textbf{High-throughput computation and machine learning
modeling of magnetic moments and Mössbauer parameters
for Fe-based intermetallics}}

\author{Bo Zhao} 
\affiliation{Institute of Materials Science, Technische Universit\"at Darmstadt, 64287 Darmstadt, Germany}

\author{Hongbin Zhang}
\email{hongbin.zhang@tu-darmstadt.de}
\affiliation{Institute of Materials Science, Technische Universit\"at Darmstadt, 64287 Darmstadt, Germany}

\begin{abstract}
    Based on high-throughput density functional theory calculations, we evaluate the local magnetic moments and Mössbauer properties for Fe-based intermetallic compounds and employ machine learning to map the local crystalline environments to such properties. It is observed that magnetic moments and Mössbauer parameters provide complementary insights into the local crystalline environment, where the statistical features cannot be captured using phenomenological models. Furthermore, we find that the scarcity of existing data in Materials Project (MP) poses a significant challenge in developing predictive machine learning models, whereas SOAP-based descriptors can be applied for reliable modeling of the enriched datasets with extra structural prototypes. This work advances the mapping of local crystalline structures to magnetic and spectroscopic properties, bridging the gap between empirical observations and theoretical models. 
\end{abstract}

\maketitle

\section{\centering Introduction}

\noindent
The local crystalline environment strongly influences the electronic structure and numerous related properties. For itinerant magnets, the presence and magnitude of a magnetic moment can be fundamentally understood through the local density of states (DOS), in accordance with the Stoner theory~\cite{kubler2000theory}. In a simplified view, a system develops from a nonmagnetic to a magnetic ground state if the product of the nonmagnetic DOS at the Fermi energy $\mathrm{N(E_F)}$ and the Stoner coefficient $I$---a constant specific to each element---exceeds unity, i.e., $I \cdot \mathrm{N(E_F)} > 1$. As a result, although isolated atoms of almost all transition metals exhibit nonzero magnetic moments, only 5 of the 30 transition elements retain magnetism in their bulk crystalline forms~\cite{gambardella2020magnetic, mohn2002magnetism}.
This phenomenon is further exemplified by dimensional crossover effects. For instance, the magnetic moment of Fe atoms in two-dimensional thin films is about $3.0\,\mu_B$~\cite{turek2006exchange}, which surpasses the $2.2\,\mu_B$ observed in body-centered-cubic (bcc) Fe. This increase can be attributed to an enhanced DOS at the Fermi energy ($\mathrm{E_F}$) stemming from a reduced coordination number, which drops from 8 in bulk bcc Fe to 5 in the thin-film configuration~\cite{Blgel:136125}. Nevertheless, it is still elusive whether such a phenomenological perspective can be applied to fully capture the complexities of structure--property relationship for a wide range of crystalline environments, where subtle details of bonding, hybridization, symmetry etc. shall be systematically considered. 

\noindent
Furthermore, M\"ossbauer spectroscopy is among the spectroscopic techniques that, by probing local electronic structures, can reveal coordination numbers~\cite{bagus2013interpretation}, spin states~\cite{henderson2014x}, magnetic moments~\cite{shehada2021trends}, and other structural features. For instance, the charge-ordering-induced Verwey transitions~\cite{verwey1939electronic} observed in magnetite compounds have been extensively investigated~\cite{senn2012charge}. In particular, precise ratios of the various iron valence states in \textit{Cc} $\mathrm{Fe_3O_4}$ below the Verwey transition temperature have been determined using $^{57}$Fe M\"ossbauer spectroscopy~\cite{vreznivcek2017understanding, krvcmavr2025charge}.
\textit{Ab initio} calculations of M\"ossbauer parameters, when benchmarked against experimental spectra, enable the prediction of a range of electronic states and offer deeper insights into the physics of specific compounds~\cite{blaha2010calculations, vreznivcek2017understanding}. Nonetheless, the limited knowledge regarding the distribution of M\"ossbauer parameters presents a persistent challenge, impeding both the accurate deconvolution of experimental spectra and the systematic benchmarking of theoretical methods.

Machine learning (ML) approaches have made rapid progress in the field of spectroscopy, enabling more effective analysis of spectral data and deeper interpretation of local crystalline environments and the resulting electronic structure~\cite{kwon2024spectroscopy}. Modern algorithms can link high-dimensional spectral signatures to subtle structural or electronic features that might otherwise go undetected. For instance, ML has been employed for predicting electron densities of states (DOS)~\cite{kong2022density, fung2022physically, how2025adaptive}. In particular, the use of Graph Neural Networks (GNNs) facilitates switching between total and partial DOS predictions, while combining probability distribution functions and cumulative distribution functions has been shown to significantly improve predictive performance~\cite{fung2022physically}.
Machine learning methods have also been utilized to map local atomic environments to X-ray spectral data. State-of-the-art models can predict core-level binding energies and spectral line shapes in X-ray photoelectron spectroscopy (XPS) with remarkable accuracy~\cite{golze2022accurate}, advancing capabilities in compositional quantification, surface chemistry analysis, and the understanding of charge-transfer phenomena. Likewise, X-ray absorption spectroscopy (XAS), known for its data-intensive nature, has benefited from ML-based inverse-problem strategies. These techniques enable the extraction of local coordination environments, oxidation states, and more nuanced electronic structure features directly from experimental spectra~\cite{carbone2024accurate, liang2023decoding}.
Multimodal strategy~\cite{jia2025revealing} has recently emerged as a useful technique to incorporate diverse spectroscopic features for comprehensive characterization.
Moreover, ML has played a key role in autonomous characterization of spectral descriptors. One example is active learning combined with Bayesian optimization for parameterizing the multiplet Hamiltonian model of NiO by sampling from the experimental XAS spectrum~\cite{zhang2023autonomous}.
With the advent of generative models such as generative adversarial networks (GANs)~\cite{long2021constrained}, variational autoencoders (VAEs)~\cite{xie2021crystal}, and diffusion models~\cite{xu2023geometric}, spectroscopy data has also begun to serve as a property within conditioned diffusion models~\cite{kwon2024spectroscopy}, expanding the possibilities for inverse design and data-driven structure--property investigations.

\noindent
However, machine learning in spectroscopy continues to face substantial challenges stemming from limited and noisy datasets. In addition to straightforward measurement errors, the dearth of physically interpretable spectral descriptors remains a primary obstacle. Typically, one aims to identify distinctive peaks, band shapes, and peak intensities that can be associated with specific molecular or electronic transitions~\cite{guda2021understanding}. When these features are not explicitly encoded, machine learning models often struggle to develop robust representations of chemical and electronic structures. Moreover, although “black-box” models such as deep neural networks can be highly effective, they afford limited interpretability if they rely solely on raw spectral data. Researchers are thus confronted with the task of balancing high predictive performance against the need for transparent, physically informed features.
M\"ossbauer spectroscopy stands out among the above-mentioned energy-scaled spectroscopy due to its direct reliance on hyperfine interactions, namely the isomer shifts, quadrupole moments, and magnetic hyperfine fields~\cite{greenwood2012mossbauer}. Machine learning using computational augmentation of these M\"ossbauer spectral descriptors can leverage underlying physical knowledge which makes it possible to trace the model back to the fundamental insights. 


In this work, we focus on the Fe-based intermetallic compounds since the magnetism of Fe depends largely on the local crystalline environments like dimensions and the coordinations. There have been extensive investigations on the composition and volume effects on various Fe-based alloy systems including ordered and disordered phases~\cite{hafner1994amorphous, khmelevskyi2004magnetostriction, wrobel2015phase}. We study the magnetic properties of iron atoms in the intermetallic compounds by looking into the magnetic moments, as well as the three M\"ossbauer parameters. 
Our work is proposed by first analyzing the trends and the dependence of magnetic moments and M\"ossbauer parameters on the local structures of Fe in Fe-X (X=B, Co, and Ti) binary systems through high-throughput (HTP) screenings, including all the possible stable and metastable structures,
along with the existing Fe-based compounds in the Materials Project~\cite{jain2013commentary} database for comparison. We further show that the four properties provide complementary information towards the complete picture of local crystalline environments, far exceeding the qualitative performance of the Stoner model. With well-defined distributions of the four properties obtained from HTP, the structure-property relationship can be well mapped by machine learning, using embedded descriptors from the local structure. 

\section{\centering Methodology}

\subsection{Data and High-throughput (HTP) Calculations}
\label{sec:HTP}
The MP datasets of existing compounds contained 2079 Fe-binary and ternary intermetallic materials with specific stoichiometric ratios and well-ordered crystal structures. Magnetic moments and M\"ossbauer parameters calculated using the methods explained below. In the following sections, we refer to this dataset as MP data for brevity. For the Fe-X (X=B, Co, and Ti) binary systems, structure relaxations were first performed to obtain optimized atomic positions and lattice constants, followed by the self-consistent calculations for magnetic moments and all the M\"ossbauer properties. The initial structures were constructed by substituting Fe and B/Co/Ti into a database containing 10493 binary structure prototypes~\cite{long2021accelerating}. The HTP screening of structure relaxations and magnetic moments was performed within the framework of the in-house developed HTP environment~\cite{singh2021multifunctional, shen2021designing, shen2023accelerated}, interfaced with the Vienna \textit{ab initio} Simulation Package (VASP). The generalized gradient approximation (GGA) of Perdew-Burke-Ernzerhof was used as the exchange-correlation functional. Relaxation was carried out in four steps with gradually increased \textit{k}-mesh and cutoff energy, from 40 to 60~\(\text{\AA}^{-3}\) and from 350 to 500~eV, respectively. Magnetic moments were obtained from the last self-consistent step, and only ferromagnetic configurations were considered under the assumption that the magnitudes of magnetic moments were independent of the magnetic ground states.

HTP calculations for Mössbauer parameters, including isomer shift (\(\mathrm{\delta_{\mathrm{IS}}}\)), quadrupole splitting (QS), and magnetic hyperfine fields (\(\mathrm{B_{hf}}\)), were carried out using WIEN2k~\cite{blaha2001wien2k}. The \emph{k}-mesh was set as \(V \cdot \mathrm{N(kpoint)} = 20,000\), where \(V\) was the volume of the cell in \(\text{\AA}^{-3}\). The average muffin-tin radius of Fe was 2.1~a.u. RKmax and GMax were set to 8.0 and 14.0 based on convergence tests. The convergence criteria for energy and charge were set to \(10^{-6}\)~Ry and \(10^{-4}\), respectively.

The isomer shift (\(\delta_{\mathrm{IS}}\)) is caused by the difference in electrostatic interaction between the potential of the nuclear charge and the electron charge density at the location of the nucleus. Influenced by the 3\(d\)-binding electrons, the effective radius for the density of \(s\)-electrons experience a change \(\Delta R\), and the resulting \(\delta_{\mathrm{IS}}\) measured experimentally is obtained by
\begin{equation}
    \delta_{\mathrm{IS}} = \frac{1}{4\pi\varepsilon_0} \frac{Ze^2}{R} \, 4\pi R^2\Delta R
    (|\psi_A|^2 - |\psi_Q|^2),
\end{equation}
where \(\psi_A\) and \(\psi_Q\) are the normalized wave functions of the \(s\)-electron for the absorber and the source.

The magnitude of the quadrupole splitting is determined by the \(z\)-component of the electric field gradient (EFG). For a nucleus with \(I = 3/2\) excited states, QS is given by
\begin{equation}
\label{eq:QS}
    QS = \frac{1}{2} e^2 q Q \bigl(1 + \eta^2/3\bigr)^{1/2},
\end{equation}
where \(eQ\) is the quadrupole moment of the nucleus, \(eq = V_{zz}\) is the \(z\)-component of the EFG, and \(\eta\) is the asymmetric parameter. The EFG, a traceless tensor of rank 2, can be obtained by DFT calculations as
\begin{equation}
    V_{ij} = \frac{\partial^2V(\vec{r})}{\partial x_i \partial x_j}\Big|_{x=0}.
\end{equation}
The principal component under Cartesian coordinates \(V_{zz}\) is obtained from the charge density \(\rho(\vec{r})\) in the following way:
\begin{equation}
    V_{zz} = \int \rho (r)\, \frac{2 P_2(\cos \theta)}{r^3} \, dr,
\end{equation}
which can be parameterized as contributions from wave functions with different angular momentum (p--p, d--d, etc.)~\cite{blaha1988first}. Instead of analyzing the entire tensor, we focus only on \(V_{zz}\) as the electric field gradient (EFG) in this study.

The magnetic hyperfine field \(B_{hf}\) can be nonzero for magnetic materials and is composed of three parts,
\(B_{hf} = B_c + B_{dip} + B_{orb}\), where \(B_c\) is the Fermi-contact field, \(B_{dip}\) is the spin-dipolar field, and \(B_{orb}\) is the orbital field~\cite{blaha2010calculations}. \(B_c\) is the dominant term related to the spin density difference at the nucleus, with the direction aligned with the nuclear moment and given by
\begin{equation}
    B_c = -\frac{e\hbar}{3mc}\langle \psi\vert \frac{\vec{\sigma}}{r^2}\delta(r)\vert \psi\rangle
    = -\frac{4\pi e\hbar}{3mc}\langle \psi\vert \vec{\sigma}\,\delta^3(r)\vert \psi\rangle,
\end{equation}
where \(\vec{\sigma}\) is the Pauli matrix and \(\delta(r)\) the delta function. $\mathrm{B_{orb}}$ and $\mathrm{B_{dip}}$ originated from orbital moments and magnetic dipole moments, i.e.,
\begin{align}
    \label{eq:orb}
    B_{orb} &= 2\mu_B \langle \psi \vert \frac{S(r)}{r^3}\,\vec{l}\,\vert \psi\rangle, \\
    \label{eq:dip}
    B_{dip} &= 2\mu_B \langle \psi\vert \frac{S(r)}{r^3} \bigl[ 3(\vec{s}\cdot\vec{r})\, \vec{r} - \vec{s}\bigr]\vert \psi\rangle,
\end{align}
where \(\mu_B\) is the Bohr magneton, and \(S(r)\) is a radial function. Although \(B_{orb}\) and \(B_{dip}\) are typically orders of magnitude smaller than \(B_c\), they can become comparable in systems with large orbital or dipolar contributions.

We validated our Mössbauer parameter calculations (isomer shift \(\delta_{\mathrm{IS}}\), electric field gradient EFG, and magnetic hyperfine fields \(B_{\mathrm{hf}}\)) by benchmarking against experimental data. As shown in Figure~S1, the results for \(\delta_{\mathrm{IS}}\) shown in panel~(a) display a systematic overestimation but are otherwise able to reproduce the overall experimental trends. EFG and $\mathrm{B_{hf}}$ are able to be calculated accurately with only few deviated points, as shown in Figure~S1(b) and (c).
This assures us that our computational approach, including exchange-correlation functionals and all-electron methods used, is capable of capturing the essential physics of the hyperfine interactions in the M\"ossbauer experiments.

\subsection{Descriptors}
In our machine learning approach, we used SOAP (smooth overlap of atomic positions)~\cite{bartok2013representing} to represent the local structures, and employed Magpie (materials-agnostic platform for informatics and exploration)~\cite{ward2016general} to represent the chemical information. In SOAP, atomic positions are encoded as smooth Gaussian distributions centered on each atom of a given species \(Z\),
\begin{equation}
\rho^Z(r) = \sum_{i}^{Z} \exp\Bigl(-\tfrac{1}{2}\sigma^2 \lvert r - R_i \rvert^2\Bigr).
\end{equation}
The distributions are then projected onto a basis of radial functions \(g_n(r)\) and spherical harmonics \(Y_{lm}(\theta, \phi)\),
\begin{equation}
\rho^Z(r) = \sum_{nlm} c^Z_{nlm} g_n(r)\,Y_{lm}(\theta, \phi).
\end{equation}
The final SOAP descriptors are formed by the power spectrum of the coefficients \(c^Z_{nlm}\), with each element given by
\begin{equation}
    P^{Z_1Z_2}_{nn'l} =
    \pi \sqrt{\tfrac{8}{2l+1}} \sum_{m} c^{Z_1}_{nlm} \, c^{Z_2}_{n'lm}.
\end{equation}

This process ensures rotational invariance, so only iron atoms with unique local environments were considered; i.e., those with duplicate Wyckoff positions were excluded from the dataset. We chose a cutoff radius \(r_{cut} = 6\)\,\AA, \(n=6\) radial basis functions, and a spherical harmonics degree \(l=6\). The dimension of SOAP also depends on the number of species in the compound. For systems involving varying numbers of species, we applied the compression method introduced by Darby et al.~\cite{darby2022compressing} to encode descriptors into a uniform length.

Magpie descriptors are a set of features designed for machine learning applications in materials science. These descriptors capture fundamental material properties based on the composition of a material, without requiring detailed structural information. They include statistical summaries (e.g., mean, minimum, maximum, and standard deviation) of elemental properties such as atomic number, electronegativity, valence electron count, and cohesive energy. Magpie descriptors are widely used for predicting material properties and for guiding high-throughput computational screening.

\subsection{Machine Learning}
We applied the random-forest algorithm to predict magnetic moments and the three M\"ossbauer parameters ($\mathrm{\delta_{IS}}$, EFG and $\mathrm{B_{hf}}$). The datasets were split into training and test subsets in an 80\%/20\% ratio. Compared to other machine learning methods, such as neural networks, random-forest excels at detecting complex patterns, evaluating feature significance, and handling outliers with reduced sensitivity~\cite{geron2022hands}. We performed tenfold cross-validation to calculate averaged predictions for each dataset. Fundamentally, the random-forest approach builds an ensemble of decision trees, each trained on a random subset of the data and considers a random subset of features during node splitting. For regression tasks, the algorithm improves prediction accuracy by averaging tree outputs (majority voting in classification tasks), thus reducing variance. Moreover, random-forest provides estimation of feature importance, helping to identify the most influential input variables that govern the model’s output. These characteristics make the algorithm not only a robust predictive tool but also a valuable means of gaining insight into key factors driving the observed trends.

\section{\centering Results and discussions}

\subsection{High-throughput (HTP) Calculations}

\begin{figure}[H]
  \centering
  \includegraphics[width=0.88\textwidth]{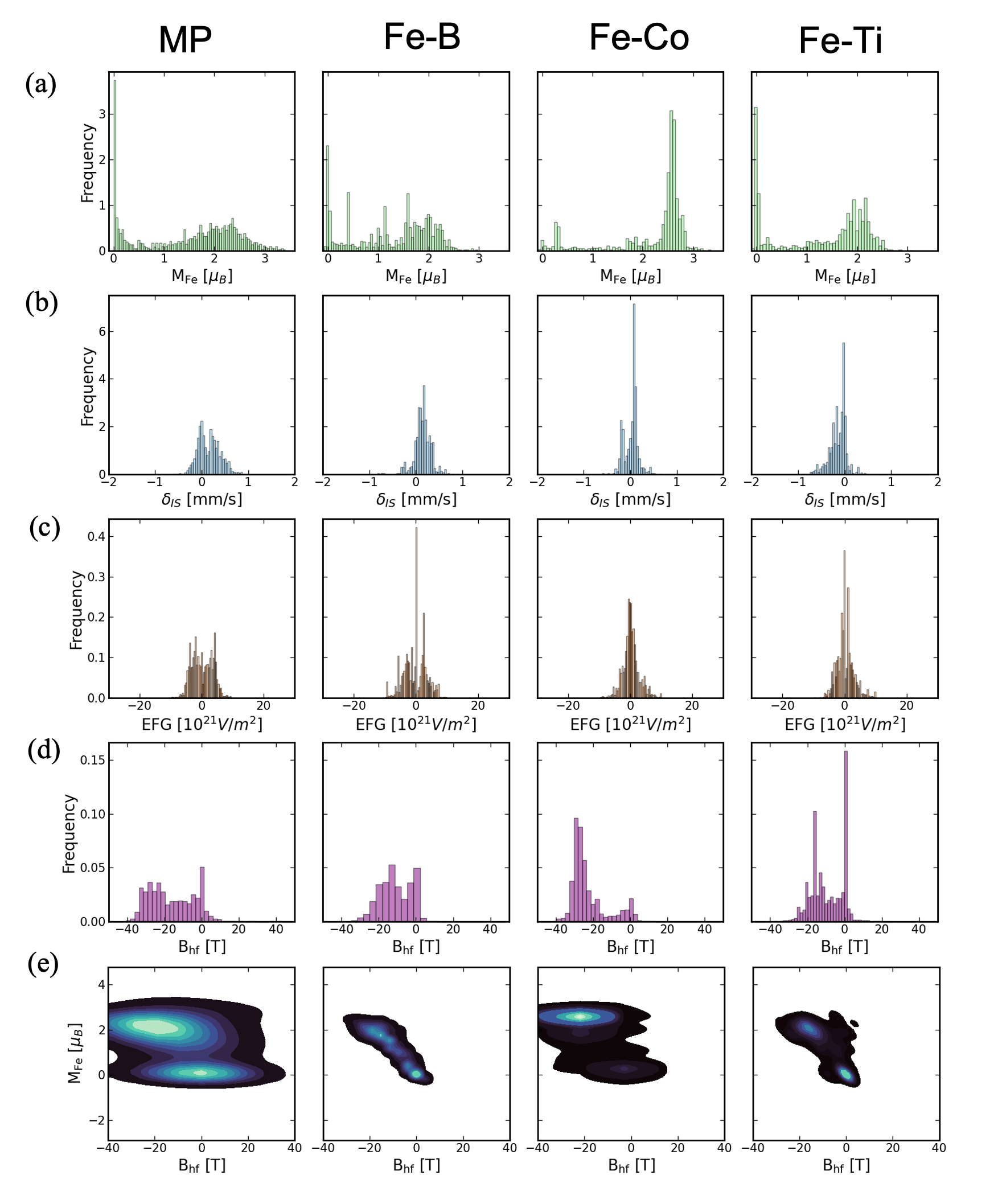}
  \caption{The distributions (normalized for better visualization) are presented as follows: (a) local magnetic moments, (b) the isomer shift, (c) the electric field gradient, and (d) the magnetic hyperfine field, along with (e) joint distribution heat maps for iron atoms in the three Fe-based binary systems, benchmarked against the Materials Project (MP) data (light color indicates higher density). }
  \label{fig:HTP_all}  
\end{figure}

Figure~\ref{fig:HTP_all}(a)-(d) displays the results of magnetic moments ($\mathrm{M_{Fe}}$), isomer shift ($\delta_{IS}$), EFG, and hyperfine fields ($\mathrm{B_{hf}}$) obtained by our HTP calculations for the Fe-B, Fe-Co, and Fe-Ti binary systems, as well as compounds from the MP database.
Based on the HTP calculations described in Sec.~\ref{sec:HTP}, we obtain 9,207, 10,177, and 10,241 converged structures with 11,354, 18,666, and 11,998 distinct Wyckoff positions for the Fe-B, Fe-Co, and Fe-Ti systems, respectively. 2154 Fe-based binary and ternary intermetallic compounds, including 3545 distinct Wyckoff positions, are collected from Materials Project (MP) datasets. The distribution patterns of these four physical properties exhibit distinct behaviors throughout the four datasets.
The magnitude of the magnetic moments stretches between 0 and 4 $\mu_B$ per Fe atom. In addition, there is a noticeable peak around 2.2~$\mu_B$ for MP data, comparable to that of bcc Fe. 
In Fe-B and Fe-Ti systems, a slight shift of the peak towards smaller value is observed. The distributions of MP, Fe-B and Fe-Ti systems indicate the robustness of Fe magnetic moment in bulk environments.
In Fe-Co system, the magnetic moments are enhanced with the peak of the histogram shifted to 2.6~$\mu_B$ as shown in Fig.~\ref{fig:HTP_all}(a), which can be ascribed to the increased or complete filling of the spin-up bands relative to those present in body-centered cubic (bcc) iron (Figure.~S3).
Similar trends are observed in the hyperfine fields ($\mathrm{B_{hf}}$), such that the average $\mathrm{B_{hf}}$ in Fe-Co is larger than that in Fe-B and Fe-Ti.
It is further observed from the correlations shown in Figure~\ref{fig:HTP_all}(e) that in Fe-B and Fe-Ti systems, the magnitude of the hyperfine field generally correlates with the magnitude of the local magnetic moments~\cite{novak2003self}, where in MP and Fe-Co systems such correlations are less obvious. The correlations also show that a small fraction of iron atoms exhibit zero $\mathrm{B_{hf}}$ despite having finite magnetic moments. This observation reflects the compensating orbital and spin-dipole contributions ($\mathrm{B_{orb}}$ and $\mathrm{B_{dip}}$) in the total hyperfine field given by Eq.~\ref{eq:orb} and~\ref{eq:dip}, which can adopt negative projections with that of the Fermi contact fields $\mathrm{B_c}$ along the magnetization axis.
The isomer shift ($\delta_{\mathrm{IS}}$) and the electric field gradient (EFG) shown in Figure~\ref{fig:HTP_all}(b) and (c) exhibit similar distributions with each others that the majority clustering around zero values.
Across the three specific iron systems, $\delta_{\mathrm{IS}}$ and EFG in Fe-Co compounds show the most centralized distribution, indicating that the electron density is close to that of $\mathrm{\alpha}$-Fe which is used as the reference. While slight shifts towards opposite directions are observed in the distributions for Fe-B and Fe-Ti, which implies a general decrease and increase of electron density, respectively. Pair-correlations for the other properties are depicted in Figure~S2. However, no obvious linear relationships are observed.
The high-throughput (HTP) calculations reveal that the four local quantities exhibit system-dependent characteristics, particularly evident in $\mathrm{M_{Fe}}$ and $\mathrm{B_{hf}}$. Aside from the parameters $\mathrm{M_{Fe}}$ and $\mathrm{B_{hf}}$, there is an absence of robust correlations between any two properties, suggesting that each property offers a distinct perspective on the underlying electronic structures. 

\begin{figure}
  \centering
  \includegraphics[width=0.88\textwidth]{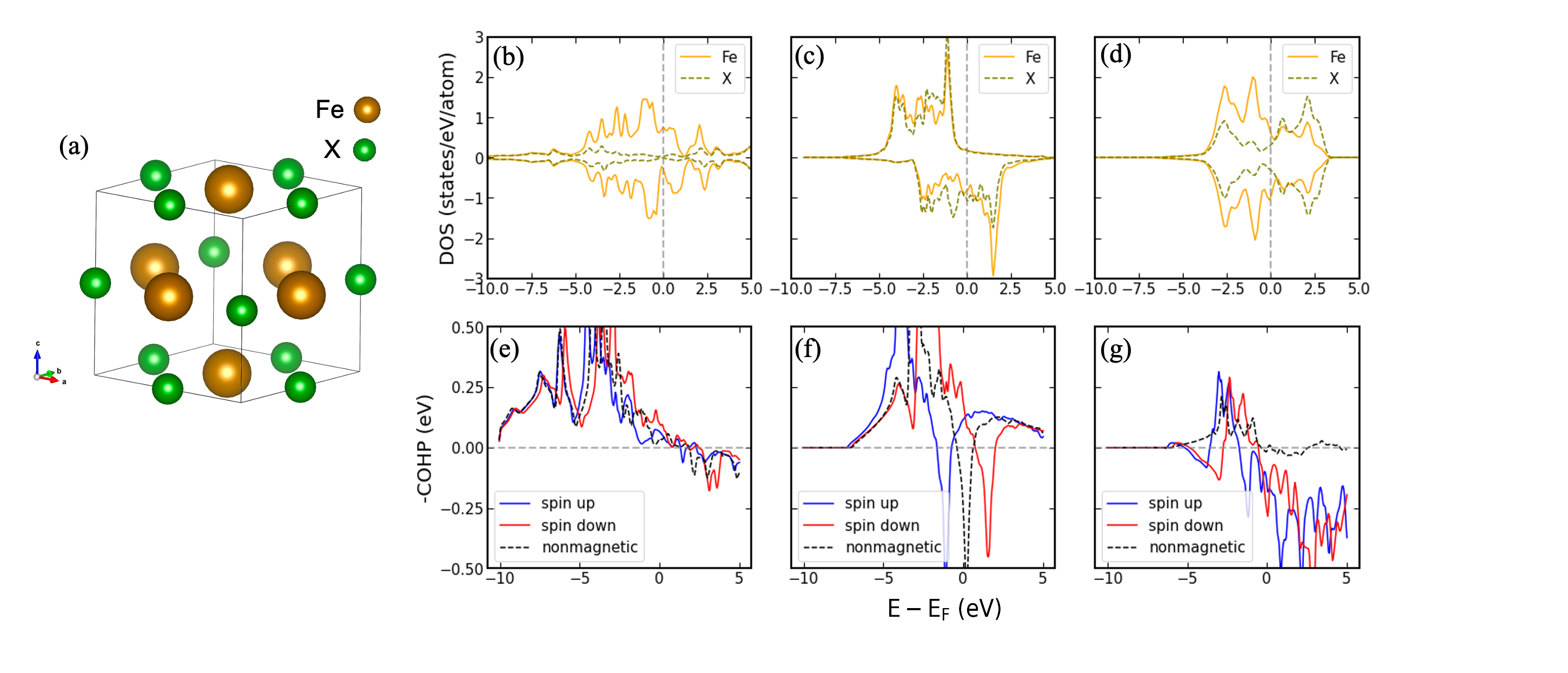}
  \caption{(a) Body center cubic ordered structure ($Pm\bar{3}m$) with Fe ratio 0.5 and the density of states (DOS) for (b) FeB, (c) FeCo, and (d) FeTi; The corresponding Fe-X crystal-orbital Hamilton population (COHP) are plotted in figure (e)-(g). Magnetic COHPs for FeB and FeTi are calculated with magnetic moments fixed at 2 $\mu_B$.}
  \label{fig:dos}  
\end{figure}
The difference in the magnitudes of magnetic moment in three Fe-X systems can be understood based on DOS and crystal-orbital Hamilton population (COHP) analyses.
Figure~\ref{fig:dos} illustrates the body-centered cubic \(Pm\bar{3}m\) FeX prototypes with iron concentration of 0.5, highlighting how the bonding and anti-bonding states differ among FeB, FeCo, and FeTi. Despite strong Fe--X hybridization in all three, FeCo attains a significantly larger value of \(2.8\,\mu_B\), while FeB and FeTi favor small or zero magnetic moments of 0.3 and 0.0 $\mu_B$, respectively. The COHP plots confirm that in FeCo, the energetic splitting of bonding vs.\ anti-bonding states in the spin-down channel localizes more electrons on Fe, promoting robust ferromagnetism. In contrast, the bonding states of Fe-X pairs in FeTi are largely reduced from magnetic to nonmagnetic configuration, confirming a more stable nonmagnetic state. FeB also stabilizes a nonmagnetic phase, evidenced by the decreased integrated -COHP (-ICOHP), corresponding to the chemical bonding strength, of Fe-X and Fe-Fe pairs by 0.09 and 0.07 eV upon spin polarization~\cite{landrum2000orbital}, as illustrated in Figure~S4.
These examples reinforce that the precise location of bonding and antibonding states with respect to the Fermi level strongly influences whether a high-spin or low-spin electronic configuration is energetically favored.

\subsection{Alignment with the Stoner Model}
\label{sec:stoner}

\begin{figure}
  \centering
  \includegraphics[width=0.88\textwidth]{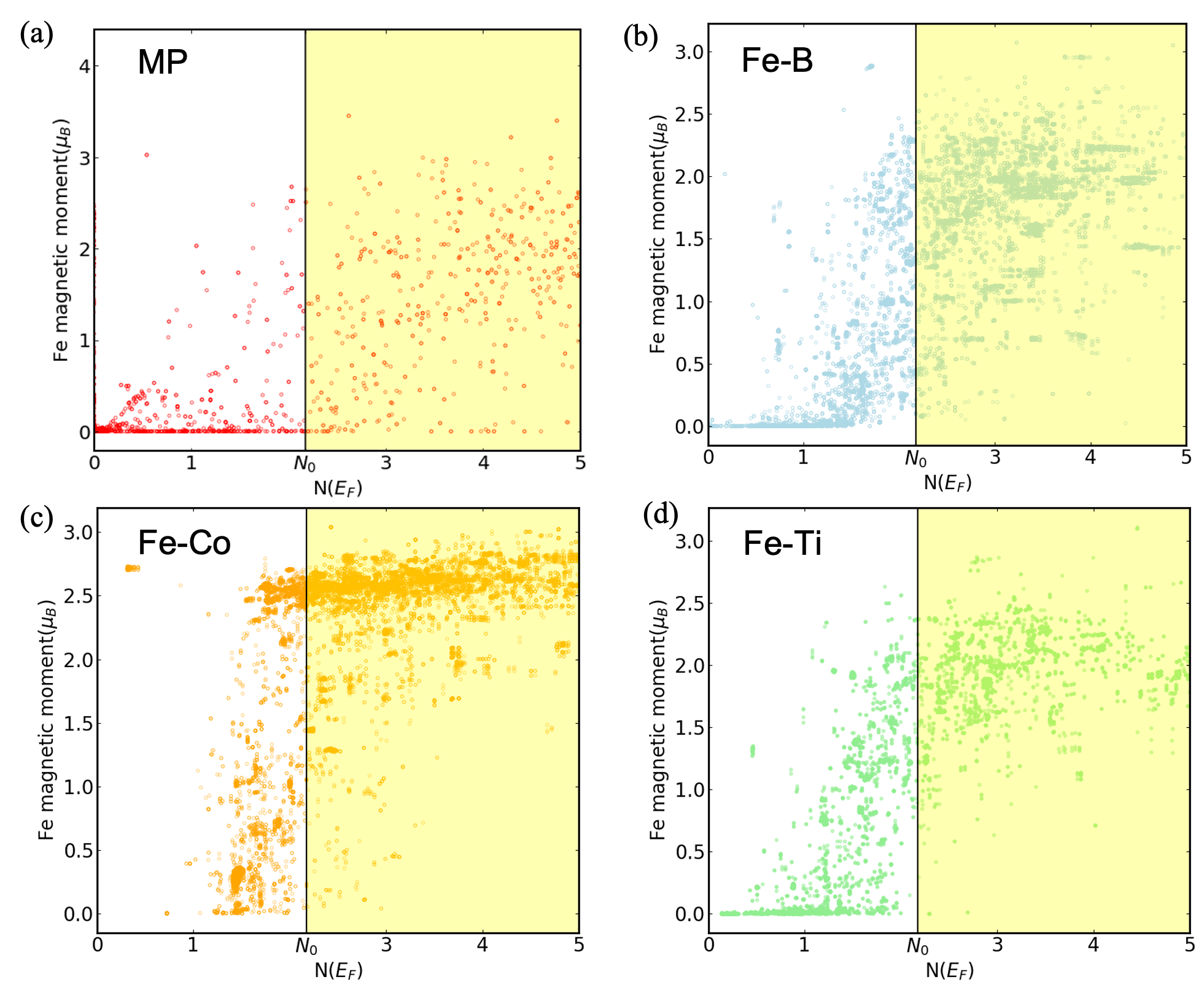}
  \caption{Fe magnetic moment vs. density of states at Fermi ($N(E_F)$) energy of non spin-polarized DOS for (a)MP, (b) Fe-B, (c) Fe-Co, and (d) Fe-Ti systems. $N_0$ is defined as the threshold $N(E_F)$ such that the $I \cdot N_0 = 1$. As suggested by Stoner theory, compounds located in the left white panel should be nonmagnetic, while those lying in the right yellow panel should be magnetic.}
  \label{fig:Stoner}  
\end{figure}

The Stoner model~\cite{stoner1939collective} has been extensively utilized for a comprehensive understanding and prediction of magnetic moment behaviors. 
According to Stoner theory, a nonmagnetic state becomes unstable relative to a ferromagnetic one if
$$
I \, n(E_F) > 1,
$$
where 
$n(E_F)$
is the non-spin-polarized density of states (DOS) at the Fermi energy, and 
$I$
is an element-specific exchange integral reflecting the strength of the exchange interaction. As shown in Figure~\ref{fig:Stoner}, we adopt 
$I = 0.46 \,\mathrm{eV}$
for Fe~\cite{janak1977uniform}, yielding a threshold 
$N_0 = 2.17$ for distinguishing between nonmagentic and magnetic Fe atoms.
Our calculations are consistent with the general trend such that Fe sites with $n(E_F) > N_0$ 
tend to develop finite magnetic moments, whereas those below the threshold generally remain nonmagnetic. However, finite magnetic moments can be obtained even with $n(E_F) < 2.17$, while nonmagnetic compounds are also among those with $n(E_F) < 2.17$ correspondingly. From Figure~\ref{fig:dos}(a) to (d), smeared onsets of $\mathrm{N(E_F)}$ distinguishing magnetic from nonmagnetic are observed for Fe-X systems, whereas less clear relationship is seen in MP system where crystalline environments are more diverse.
It is also noticeable that the magnetic moments in the four systems reach a saturation value around 3 $\mu_B$ and do not increase further with increased $n(E_F)$. 
These behaviors underscore the limitations of relying on a single parameter to model exchange effects. 
Factors such as orbital-dependent interactions, magnetic frustration, and spin fluctuations can introduce deviations from a simple rigid-band picture. Moreover, the local electronic environment—which can differ from site to site within a crystal—may shift $n(E_F)$ or alter hybridization in ways not accounted for by purely global or average DOS considerations. 

\subsection{M\"ossbauer Parameters as Additional Magnetic Characterizations}
\label{sec:hff}

\begin{figure}
  \centering
  \includegraphics[width=0.88\textwidth]{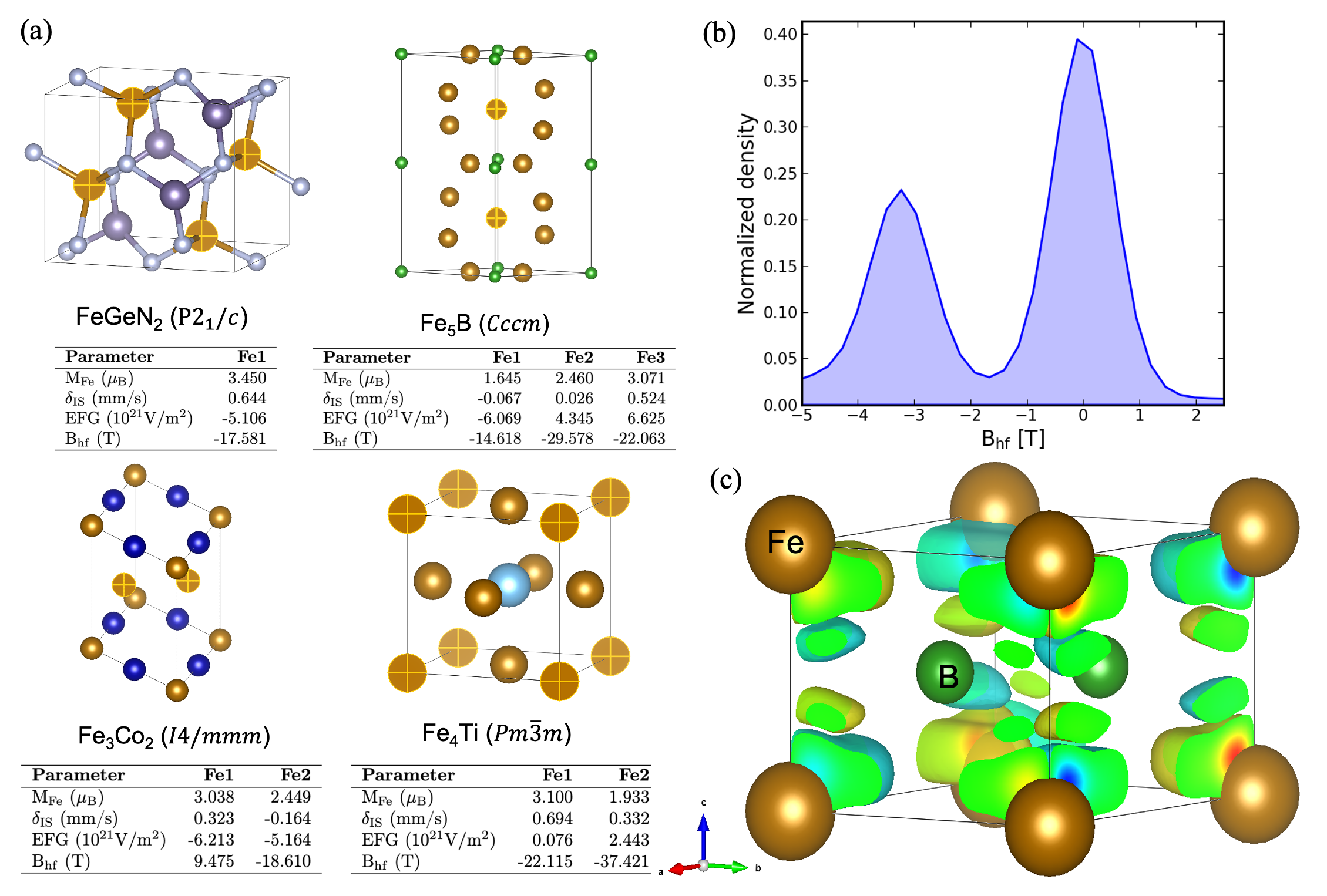}
  \caption{(a) Compounds containing Fe atoms $\mathrm{M_{Fe}} > 3 \mu_B$ and their corresponding M\"ossbauer parameters (atoms with magnetic moment larger than 3 $\mu_B$ are highlighted in the structures); (b) Normalized distribution of $\mathrm{B_{hf}}$ for nonmagnetic Fe-B compounds ($M_{Fe} < 0.5 \mu_B$); (c) Spin density of Fe in a P6/mmm hexagonal $\mathrm{FeB_2}$ compound with zero magnetic moment but large HFF value}
  \label{fig:B_HFF}  
\end{figure}

M\"ossbauer parameters, e.g., hyperfine-field ($\mathrm{B_{hf}}$), provide complementary insights into magnetic ordering and spin anisotropy, beyond what is conveyed by magnetic moments alone.
Figure~\ref{fig:B_HFF}(a) shows four compounds containing Fe atoms with magnetic moments greater than 3 $\mu_B$, along with 
their M\"ossbauer parameters. 
The compound $\mathrm{FeGeN_2}$ is observed to possess a maximum magnetic moment of 3.45 $\mu_B$, approaching the atomic magnetic limit of iron. Within this compound, the iron atoms are characterized by a notably large $\mathrm{\delta_{IS}}$ at 0.644 mm/s, and EFG at -5.106 $\mathrm{10^{21}V/m^2}$, indicating a pronounced anisotropy in the spatial distribution of the valence electrons.
For the other three compounds ($\mathrm{Fe_3B}$, $\mathrm{Fe_3Co_2}$ and $\mathrm{Fe_4Ti}$), despite similar large magnetic moments, their M\"ossbauer parameters are quite diverse, where in $\mathrm{Fe_5B}$, the large $\mathrm{M_{Fe}}$ is correlated with the large EFG at 6.625 $\mathrm{10^{21}V/m^2}$, in $\mathrm{Fe_3Co_2}$ large negative EFG (-6.213 $\mathrm{10^{21}V/m^2}$) and small $\mathrm{B_{hf}}$ (9.475 T) are found, while in $\mathrm{Fe_4Ti}$ the large $\mathrm{\delta_{IS}}$ at 0.694 mm/s may play a role.
It is also observed in Figure~\ref{fig:HTP_all} that there is a large proportion of Fe-B compounds that are nonmagnetic. However, from the $\mathrm{B_{hf}}$ distribution zooming into the nonmagnetic area (Figure~\ref{fig:B_HFF}(b)) one finds a broad distribution of $\mathrm{B_{hf}}$s and that nearly one-third of these nonmagnetic Fe atoms exhibit finite $\mathrm{B_{hf}}$ values with magnitude greater than 1 Tesla. In Figure~\ref{fig:B_HFF}(c), we show an example of a hexagonal $\mathrm{FeB_2}$ (P6/mmm) structure with nonmagnetic Fe, along with the spin density distribution. However, the Fe atom in such environment has large spin dipolar and orbital fields that are $\mathrm{B_{dip}} = 3.5$ T and $\mathrm{B_{orb}} = 1.9$ T, which can be ascribed to the ashpericity of the spin and charge density as depicted in Figure~\ref{fig:B_HFF}(b).
In conclusion, the large variation in Fe magnetic moments arises from a confluence of structural, chemical, and electronic factors, showing that no straightforward correlation with the three M\"ossbauer parameters can be definitively established and thereby requiring machine-learning approaches to reveal the underlying complexity. Furthermore, our example of $\mathrm{B_{hf}}$ in nonmagentic FeB illustrates how Mössbauer spectroscopy can capture subtle magneto-chemical environments—exposing local spin anisotropy and orbital magnetization that would otherwise remain unnoticed if only magnetic moments were examined.

Given the inherent difficulties in applying phenomenological theories, such as the Stoner model, to accurately characterize the statistical tendencies of magnetic moments, combined with the significant discrepancies observed in the distributions of the four physical properties analyzed, the pivotal question arises of how one can establish a reliable mapping between the local crystalline environments and these specific physical properties. To initially assess the capabilities of the machine learning model, a straightforward classification task is conducted distinguishing between magnetic and nonmagnetic materials. This was achieved by applying various threshold values ranging from 0.1 to 1 $\mu_B$, utilizing SOAP as the descriptor tool. The experiment demonstrates that the model achieves an accuracy exceeding 90\% across all four datasets tested. Notably, accuracy peaks at 99\% within the three Fe-X datasets, as illustrated in Figure~S5. This performance surpasses that of the Stoner model, which uses electron density at the Fermi level as its evaluative metric.

\subsection{MP Data Exploration and Evaluation}
Compared with the other Fe-X datasets, the existing Materials Project (MP) dataset, despite its extensive scope, requires significant improvement to enhance its utility in describing local magnetic environments. 
To reduces the influence of ambiguous or skewed data and to yield a more consistent and reliable distribution of magnetic properties, we focus on magnetic data with \(\mathrm{M_{Fe}} > 0.5 \, \mu_B\) and perform regression tasks to further refine the predictions of local magnetic moments. 

\begin{figure}[H]
  \centering
  \includegraphics[width=0.88\textwidth]{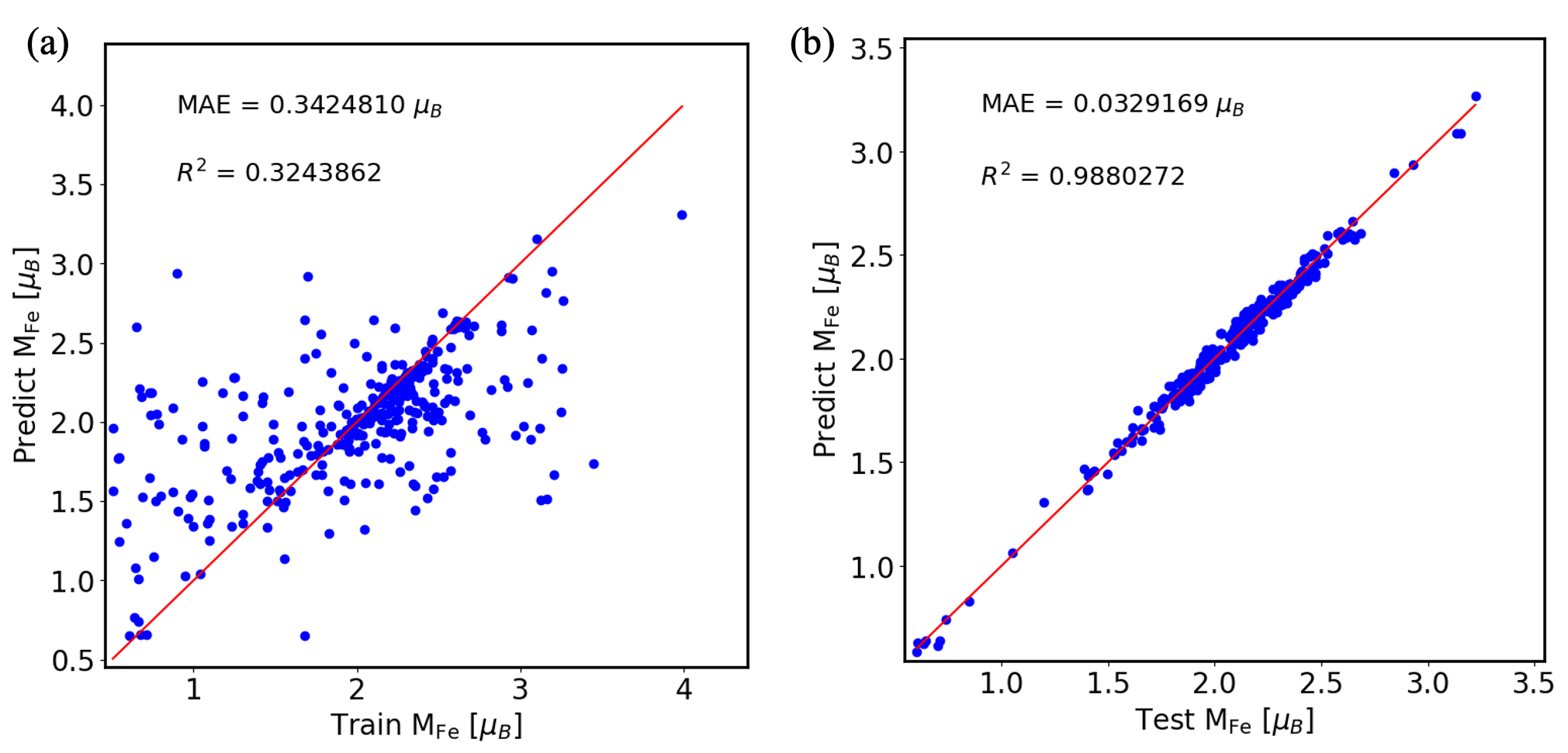}
  \caption{(a) Direct random forest regression for $\mathrm{M_{Fe}}$ on MP dataset; and (b) Random forest regression with active sampling.}
  \label{fig:MP-AL}  
\end{figure}

Using local structure-embedded SOAP descriptors, a direct random-forest regression on 2,358 iron local environments within the MP dataset achieves moderate accuracy (\(R^2 \approx 0.42\)) for predicting magnetic moments (Figure.~\ref{fig:MP-AL}(a)). This limitation is partly attributed to the sparsity of the data. This can be confirmed by manually selection of representative data into training set. For instance, by starting with an initial sample of 50 randomly selected data points and iteratively enriching the datasets with the 50 most poorly predicted points, the regression \(R^2\) rises to 0.99 for the top 20\% best predictable samples (Figure.~\ref{fig:MP-AL}(b)).
These findings highlight the necessity of targeted refinement to address the bias and the heterogeneity within the MP datasets. Augmenting its entries with strategically guided DFT calculations can significantly enhance its quality and predictive capacity, as demonstrated for the Fe-X binary systems below. 

\subsection{Machine Learning Predictions for Magnetization and M\"ossbauer Parameters}



In predicting the local magnetic moments and the M\"ossbauer parameters, physically both chemical information and structural information are needed to construct the descriptors. For the structure diverse Fe-X systems, chemical information is largely limited by the constant composition, while structural information alone is able to yield comparable accuracy with that using combined information. 
The impact of different feature representations, Magpie (global composition), SOAP (local atomic environments), and Magpie+SOAP, on the prediction accuracy for $\mathrm{M_{Fe}}$ and $\mathrm{B_{hf}}$ is assessed. For cases using Magpie descriptors only, compound-wise global $\mathrm{M_{Fe}}$ and $\mathrm{B_{hf}}$ are used as properties, which are obtained by the weighted sum of the corresponding local quantities with weights given by the atomic fraction. Magpie encodes aggregate composition-based features (e.g., overall stoichiometry, average electronegativity), which can capture broad chemical trends but miss site-specific details. SOAP, on the other hand, focuses on the local environment around each atom (bond lengths, coordination geometry), making them well suited for capturing subtle variations in electronic structure that drive local properties such as magnetization and hyperfine parameters.
From the results in Table~\ref{tab:ML_results}, it shows that Magpie performs much better in the prediction of global properties on MP datasets which have more varieties of chemical compositions. An exception found in $\mathrm{B_{hf}}$ property on Fe-Ti datasets, with Magpie performs comparably well. The synergy of using combined (Magpie+SOAP) descriptors also have effects to some extends only on these datasets and properties. 
The difference behavior of Magpie in various datasets can be explained by the feature importance analysis (Fig.~S6) showing that \textit{var\_EffectiveCoordination} (variation of elemental coordination numbers) and \textit{mean\_GSmagmom} (average elemental ground-state magnetic moment) are the most prominent features respectively for $\mathrm{M_{Fe}}$ and $\mathrm{B_{hf}}$ in MP datasets, while all features contribute subtly with importance smaller than 0.1 in the other system specific datasets. It also implies the fact that other than magnetic moment that depends mostly on stoichiometry, $\mathrm{B_{hf}}$ is more correlated with electronic structures, specifically the magnetic ordering approximated by \textit{mean\_GSmagmom}, which is consistent with the smeared dependency between $\mathrm{M_{Fe}}$ and $\mathrm{B_{hf}}$ as observed in Figure~\ref{fig:HTP_all}(e). Teng et al.~\cite{long2021accelerating} have also found that SOAP significantly enhances the prediction of $T_C$ among the isomer intermetallic compounds (i.e. compounds with same composition but different crystal structures), aligned with our discovery of the outstanding performance of SOAP applied on the system-specific datasets.
Thus, considering the system-specific Fe-X datasets, using only structure-embedded SOAP descriptors is reasonable to expect accurate prediction of $\mathrm{M_{Fe}}$ and M\"ossbauer parameters.
\begin{table}[!ht]
\caption{\label{tab:ML_results} Machine learning results for magnetic moments ($\mathrm{M_{Fe}}$) and hyperfine fields ($\mathrm{B_{hf}}$) on four different datasets (MP, Fe-B, Fe-Co, and Fe-Ti), using Magpie only, SOAP only, and Magpie+SOAP descriptors respectively. In the Magpie column, total $\mathrm{M_{Fe}}$ and $\mathrm{B_{hf}}$ are used as the properties.}
\begin{ruledtabular}
\begin{tabular}{llccc}
Property & Dataset & Magpie & SOAP & Magpie+SOAP \\ \colrule
\multirow{4}{*}{$\mathrm{M_{Fe}}$ [$\mu_B$]} 
 & MP    & 0.792 & 0.420 & 0.492 \\
 & Fe-B  & 0.638 & 0.938 & 0.938 \\
 & Fe-Co & 0.225 & 0.910 & 0.911 \\
 & Fe-Ti & 0.426 & 0.943 & 0.943 \\ \colrule
\multirow{4}{*}{$\mathrm{B_{hf}}$ [T]} 
 & MP    & 0.785 & 0.852 & 0.877 \\
 & Fe-B  & 0.494 & 0.938 & 0.937 \\
 & Fe-Co & 0.206 & 0.959 & 0.993 \\
 & Fe-Ti & 0.894 & 0.927 & 0.979 \\
\end{tabular}
\end{ruledtabular}
\end{table}
\begin{figure}[H]
  \centering
  \includegraphics[width=0.88\textwidth]{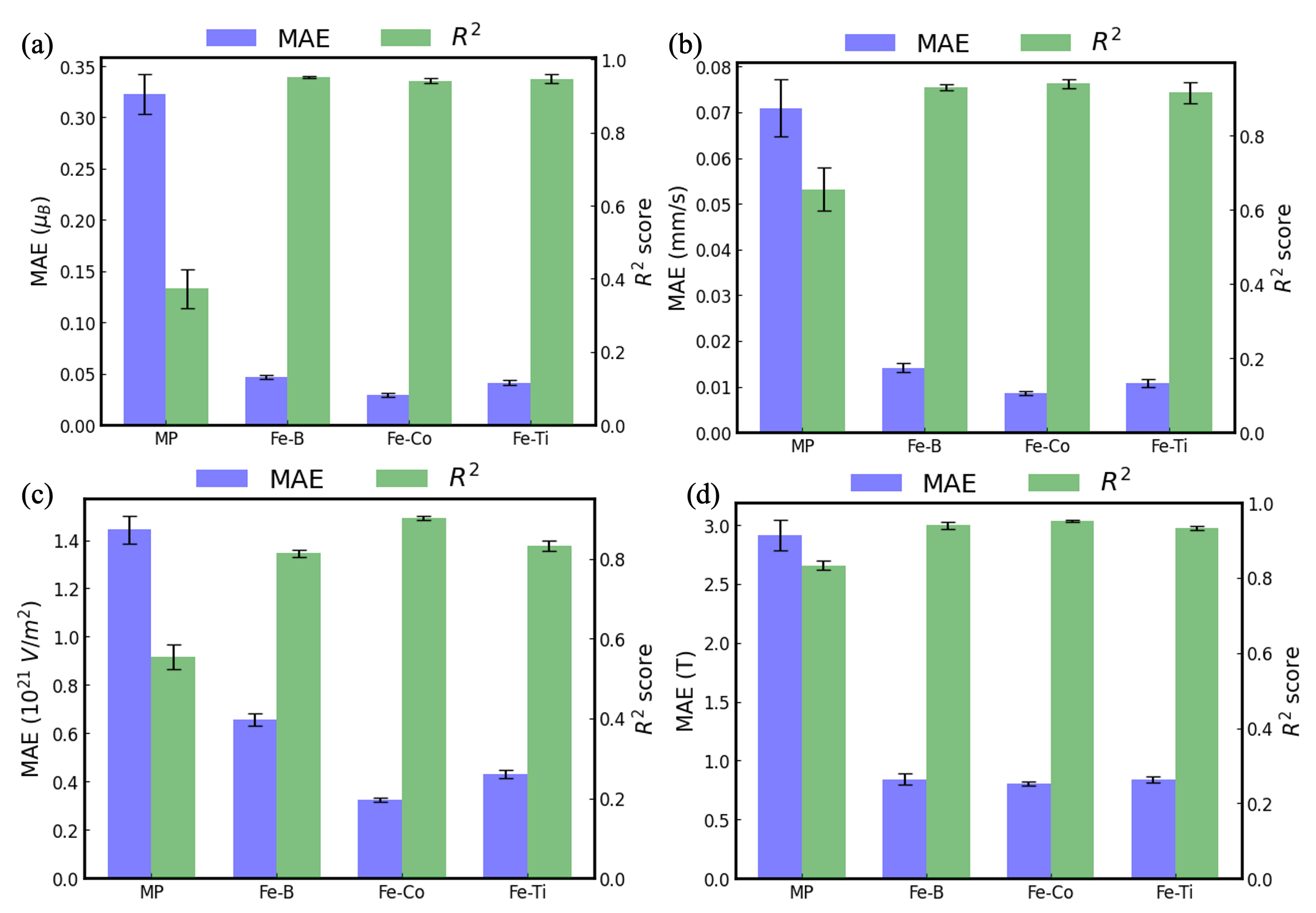}
  \caption{The prediction results of (a) magnetic moments, (b) isomer shift, (c) electric field gradient, and (d) magnetic hyperfine fields on full data of 4 different datasets, i.e., Materials Project (MP), Fe-B system, Fe-Co system, and Fe-Ti system, evaluated by the mean absolute error (MAE) and the $R^2$ score, represented by the blue and the green bars.}
  \label{fig:mag_hff_RF}  
\end{figure}

Regression of all the four local properties ($\mathrm{M_{Fe}}$,  $\mathrm{\delta_{IS}}$, EFG, and $\mathrm{B_{hf}}$) are trained with performance measured by the mean absolute error (MAE) and $R^2$ values, as depicted in Figure~\ref{fig:mag_hff_RF}.
It can be seen that the regression models achieve \(R^2\) scores exceeding 0.9 for $\mathrm{M_{Fe}}$, $\mathrm{\delta_{IS}}$, and $\mathrm{B_{hf}}$ in the Fe-B, Fe-Co, and Fe-Ti datasets. MP data remain more challenging, mainly due to broader chemical diversity and smaller, noisier sampling of local environments. Nonetheless, hyperfine fields in MP data still attain \(R^2 > 0.85\), suggesting that a structural descriptor-based mapping to $\mathrm{B_{hf}}$ is comparatively robust.
Notably, the electric field gradient (EFG) proves to be the most difficult to predict due to its high sensitivity to small distortions, local crystal fields, and site-symmetry considerations not fully captured by current descriptors. These affect not only the magnitude but also the anisotropy of the spin density as demonstrated in Sec.~\ref{sec:hff}. Schwarz et al.~\cite{schwarz1990charge,blaha1988first} claims from first-principle calculation point of view that accurate EFG requires accurate description of the partial charges, e.g., p and d valence electrons, which, however, is not able to improve in our case by simply increasing the angular momentum l used in SOAP descriptors. We have found that by increasing from l=6 to l=12 the performance in MP datasets remains unchanged (Figure.~S7). This is consistent with the current challenges in machine learning of electronic structures~\cite{kulik2022roadmap,shao2023machine}. What is also neglected is the tensor nature of EFG, while in our case a scalar quantity corresponding to the principle component of the tensor is used to represent such interaction of quadruple moments between nucleus and electrons. Future work that incorporates equivariant neural-network encoders~\cite{geiger2022e3nn, batzner20223}, which are designed to preserve and exploit local symmetry operations (rotations, reflections, translations), may provide the necessary level of geometric awareness to boost prediction accuracy for the whole EFG tensor. With both magnetic moments and different M\"ossbauer hyperfine parameters available, multimodal machine learning can be employed to synergistically strengthen the local structure-property mapping.

\section{\centering Conclusions}

In conclusion, we establish statistical mappings from the local crystalline environments to the magnetic moments and M\"ossbauer spectroscopies using machine learning. We perform high-throughput calculations on the four properties ($\mathrm{M_{Fe}}$, $\mathrm{\delta_{IS}}$, EFG, and $\mathrm{B_{hf}}$) for four intermetallic Fe-based systems that are Fe binary and ternary compounds from Materials Project (MP), and our HTP-screened Fe-B, Fe-Co, and Fe-Ti systems. The variation in the magnetic moments of Fe, apart from chemical compositions, can originate from different hybridization natures with neighboring atoms. The justification of Stoner theory on the Fe-X systems implies that the it can serve as a guidance to the magnitude of magnetic moments but the performance using such single parameter is severely limited. The inclusion of M\"ossbauer offer insights into diverse magnetic moments from different perspectives. In particular, the hyperfine field($B_{hf}$) serves as a sensitive measure of spin anisotropy. A specific example using nonmagnetic $\mathrm{FeB_2}$ demonstrates how site-specific hyperfine fields reveal subtle magnetic variations that are not always obvious from bulk magnetization alone, enriching our understanding of local spin and orbital contributions.

We find that existing MP data can be noisy or incomplete for local magnetic properties. A data cleaning by selecting representative data into training set substantially enhances predictive performance, underscoring the importance of targeted data augmentation for high-throughput materials databases. In comparison, machine-learning models trained on system-specific achieve excellent predictive accuracy for both magnetic moments and Mössbauer parameters, with $R^2$ values exceeding 0.9 in most cases. These models demonstrate the possibility and necessity to incorporate different properties for multimodal machine learning to synergistically strengthen the local structure-property mapping.


\section{Acknowledgment}
    
The authors would like to thank Prof.\ Stephan Bl\"ugel for his valuable discussions and insightful suggestions. The authors gratefully acknowledge the computing time provided to them on the high-performance computer Lichtenberg at the NHR Centers NHR4CES at TU Darmstadt. This is funded by the Federal Ministry of Education and Research, and the state governments participating on the basis of the resolutions of the GWK for national high performance computing at universities (www.nhr-verein.de/unsere-partner). This work is funded by the Deutsche Forschungsgemeinschaft (DFG, German Research Foundation) – CRC 1487, "Iron, upgraded!" – Project number 443703006.
\bibliography{references}

\section{Supplement}
\setcounter{figure}{0}
\renewcommand{\thefigure}{S\arabic{figure}}  
\setcounter{table}{0}
\renewcommand{\thetable}{S\arabic{table}}


\begin{figure}[H]
  \centering
  \includegraphics[width=0.85\textwidth]{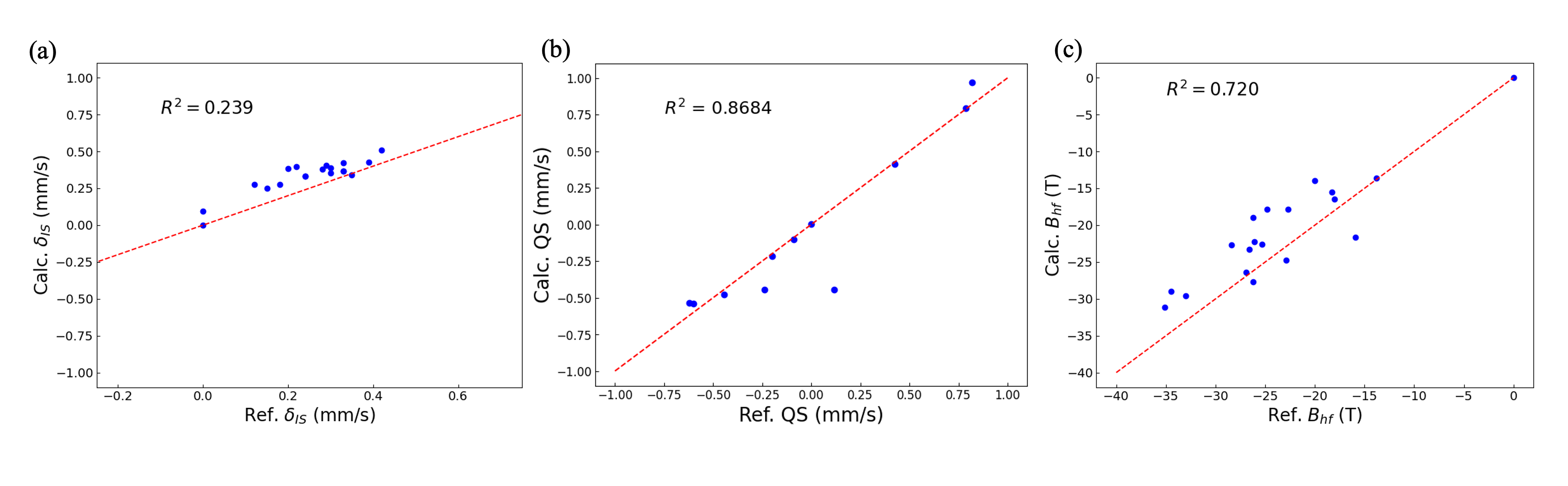}
  \caption{Comparison between calculated and experimental (a) isomer shift, (b) quadruple splitting (QS) calculated using Eq.~\ref{eq:QS}, and (c) magnetic hyperfine field of Fe in intermetallic compounds. Red dashed lines are the parity lines.}
  \label{fig:exp_calc}
\end{figure}

\clearpage
\begin{table}[h!]
\centering
\caption{List of experimental M\"ossbauer parameters for the Fe-based intermetallic compounds used in the validation in Fig.~S\ref{fig:exp_calc}}
\begin{ruledtabular}
\begin{tabular}{llll}
\textbf{Compound} & \textbf{$\delta_{IS}$ (mm/s)} & \textbf{QS} (mm/s) & \textbf{$B_{hf}$ (T)}\\
\midrule
\texttt{$\mathrm{bccFe}$}  & 0.0 & 0.0 & -33\\
\texttt{$\gamma$-FeC} & 0.42 & 0.0 & 0.0 \\
\texttt{$\eta-\mathrm{Fe_2C}$} & 0.29 & -0.04 & -18\\
\texttt{$c-\mathrm{Fe_2C}$} & 0.3 & -0.04 & -18.3 \\
\texttt{$\mathrm{Fe_3C}$ 1} & 0.33 & 0.0 & -25.3\\
\texttt{$\mathrm{Fe_3C}$ 2} & 0.33 & 0.03 & -26.6 \\
\texttt{$\gamma'-\mathrm{Fe_4C}$ 1} & 0.28 & 0.0 & -34.5 \\
\texttt{$\gamma'-\mathrm{Fe_4C}$ 2} & 0.24 & -0.09 & -26.2 \\
\texttt{$\gamma''-\mathrm{Fe_4C}$}& 0.12 & -0.1 &  -22.9 \\
\texttt{$\mathrm{Fe_5C_2}$ 1} & 0.39 & 0.12 & -26.1\\
\texttt{$\mathrm{Fe_5C_2}$ 2} & 0.25 & 0 & -22.7 \\
\texttt{$h-\mathrm{Fe_7C_3}$ 1} & 0.3 & -0.06 & -24.8 \\
\texttt{$h-\mathrm{Fe_7C_3}$ 2} & 0.22 & 0.04 & -15.9 \\
\texttt{$\alpha-Fe_{16}C_2$ 1}  & 0.0 & -0.05 & -26.9\\
\texttt{$\alpha-Fe_{16}C_2$ 2} & 0.15 & -0.04 & -28.4 \\
\texttt{$\alpha-Fe_{16}C_2$ 3} & 0.18 & -0.01 & -35.1 \\
\texttt{$\mathrm{Fe_2P}$ 1} & & -0.2 & \\
\texttt{$\mathrm{Fe_2P}$ 2} & & -0.09 & \\
\texttt{$\mathrm{Fe_4N}$ 1} & 0.329 & -0.24 & -21.7 \\
\texttt{$\mathrm{Fe_4N}$ 2} & 0.25 & 0.0 & -34.1 \\
\texttt{$\mathrm{Fe_2S}$(pyrit)} & & -0.621 & \\
\texttt{$\mathrm{Fe_2S}$(marcasite)} & & -0.6 & \\
\texttt{$\mathrm{FeNi}$} & & 0.426 & \\
\texttt{$\mathrm{FeSi}$} & & 0.787 & \\
\texttt{$\mathrm{Fe_2Y}$} & & -0.44 & \\
\texttt{$\mathrm{FeZr_3}$} & & 0.82 & \\
\end{tabular}
\end{ruledtabular}
\label{tab:iron_carbides}
\end{table}

\begin{figure}[H]
  \centering
  \includegraphics[width=0.96\textwidth]{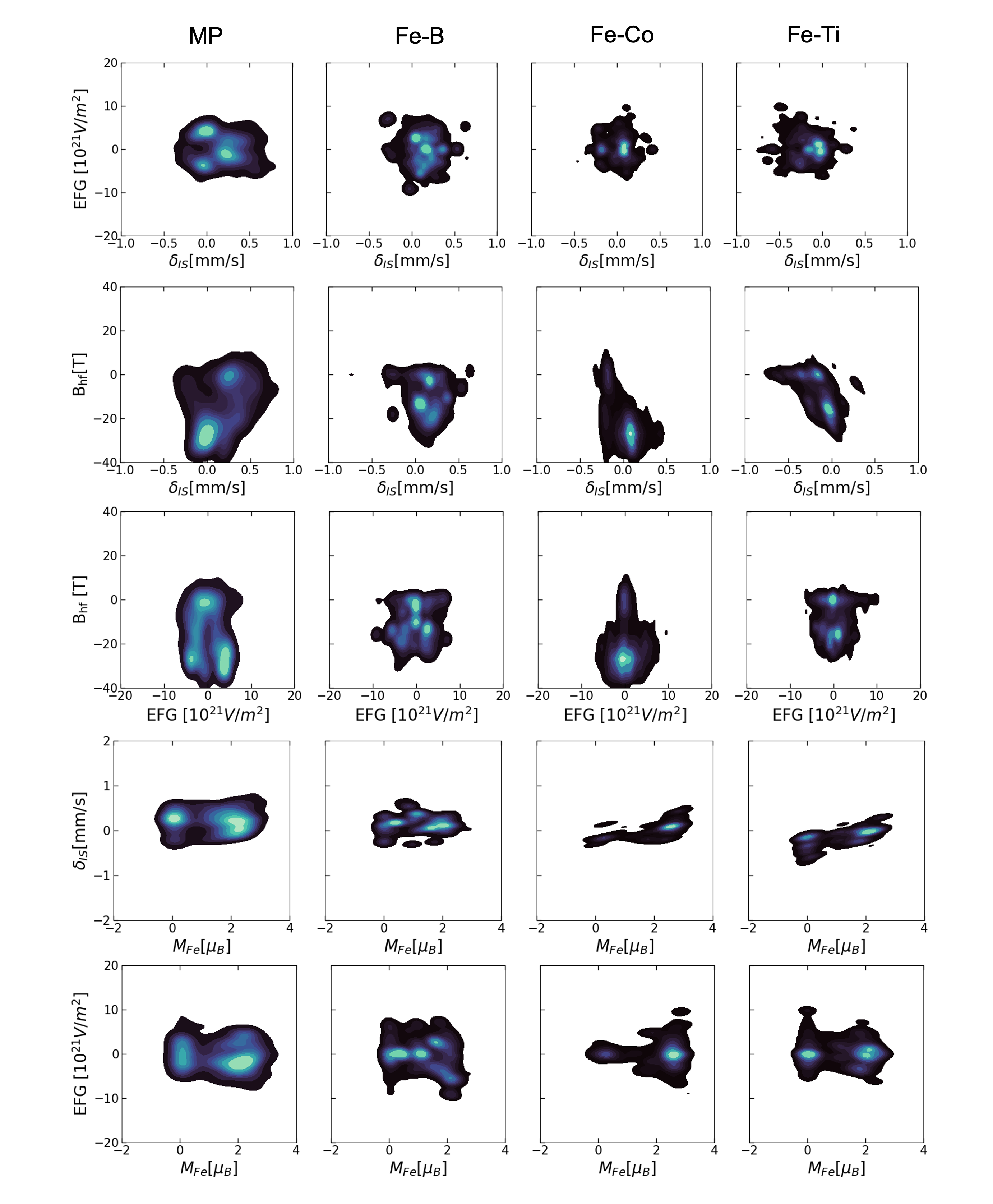}
  \caption{Heat maps of joint distribution for the four datasets between the properties (from top to bottom): $\mathrm{\delta_{IS}}$ - EFG, $\mathrm{\delta_{IS}}$ - $\mathrm{B_{hf}}$, EFG - $\mathrm{B_{hf}}$, $\mathrm{M_{Fe}}$ - $\mathrm{\delta_{IS}}$, and $\mathrm{M_{Fe}}$ - EFG. }
  \label{fig:corr_supp}
\end{figure}

\begin{figure}[H]
  \centering
  \includegraphics[width=0.96\textwidth]{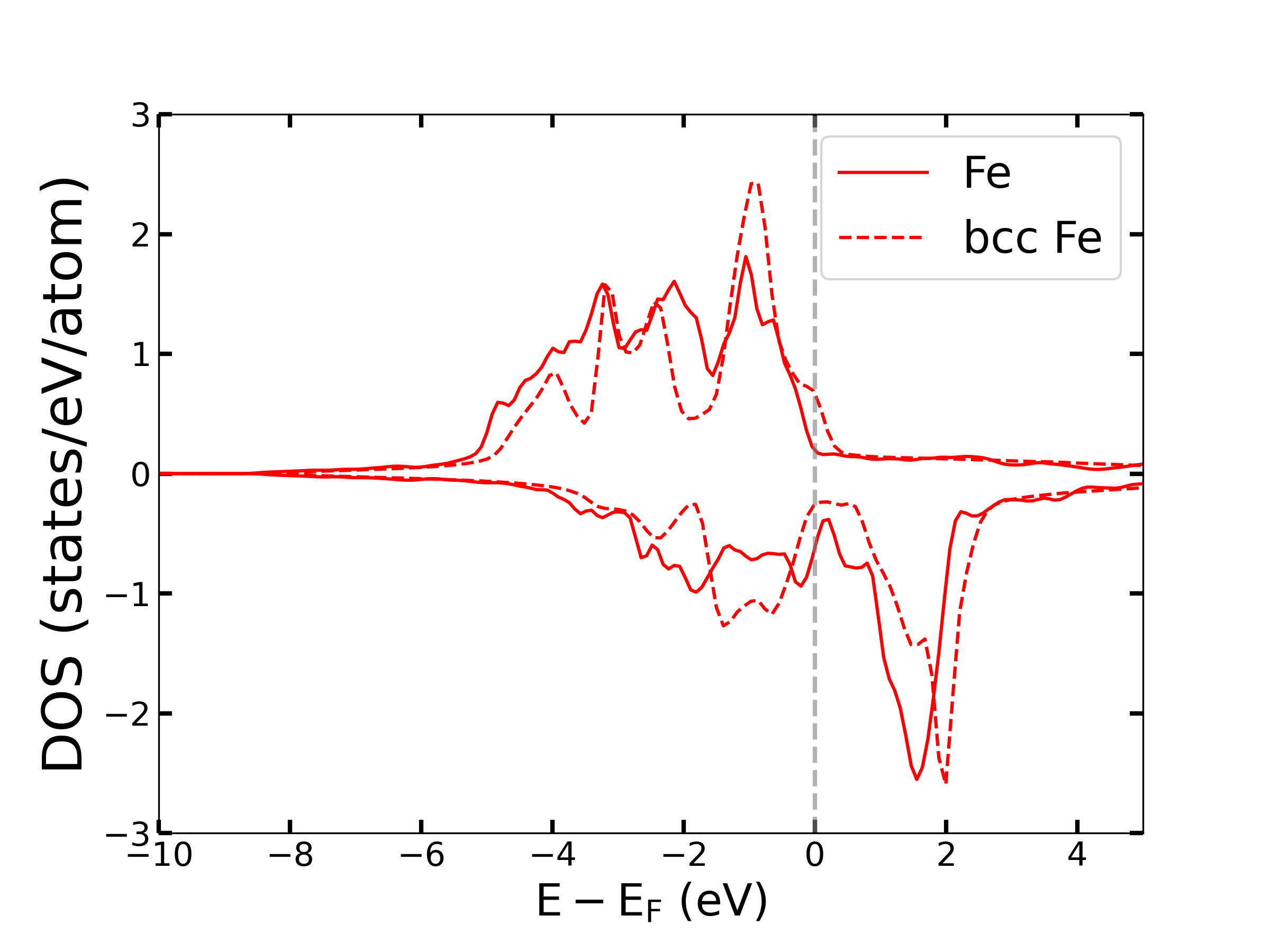}
  \caption{Spin-polarized density of states for Fe atoms in $\mathrm{Fe_5Co_9}$ (solid lines) and bcc Fe (dashed lines). In bcc Fe, there is a small fraction of unoccupied electrons in the majority spin channel, while in $\mathrm{Fe_5Co_9}$ the majority bands are fully occupied.}
  \label{fig:FeCo_Fe_dos}
\end{figure}

\begin{figure}[H]
  \centering
\includegraphics[width=0.67\textwidth]{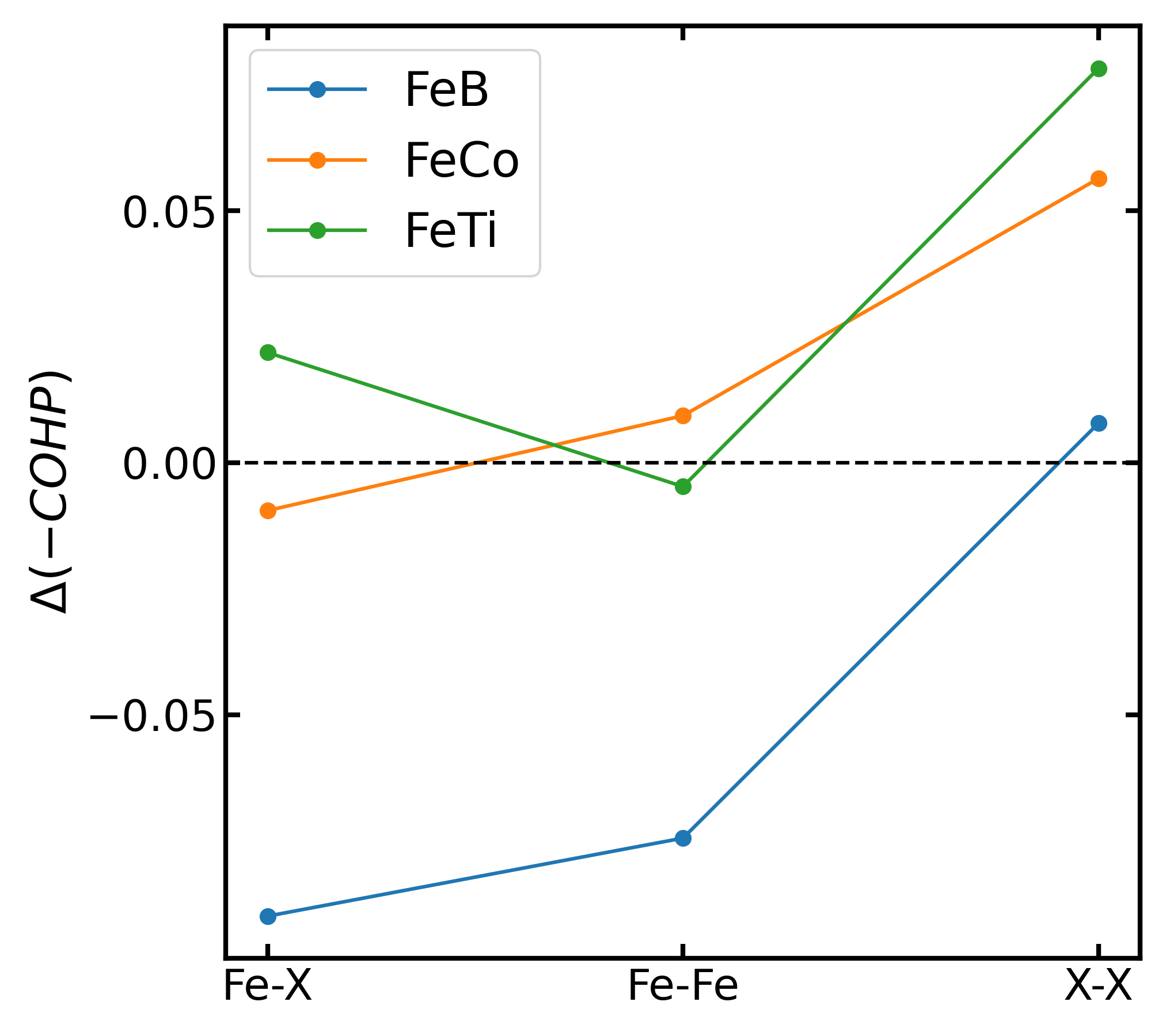}
  \caption{Change of -ICOHP ($\Delta$-COHP) for 
the Fe-X, Fe-Fe and X-X pairs upon spin polarization,i.e., $\mathrm{\Delta-ICOHP} = (\mathrm{-ICOPH_{mag})-(\mathrm{-ICOPH_{mag}})}$. Positive $\Delta$-COHP indicates bond strength enhanced and negative values correspond to the wekening of bond strength.}
  \label{fig:ICOHP}
\end{figure}

\begin{figure}[H]
  \centering
  \includegraphics[width=0.96\textwidth]{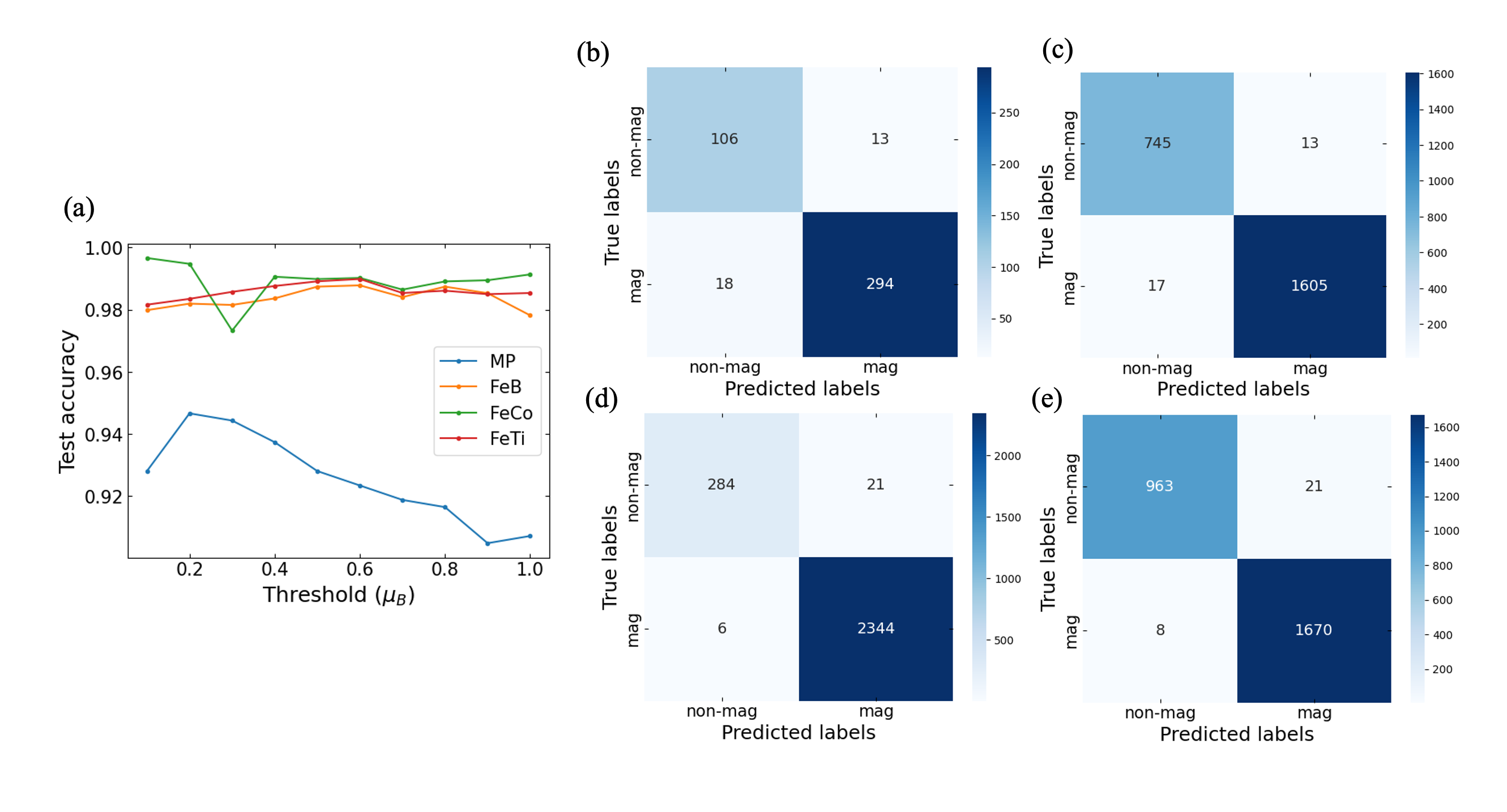}
  \caption{(a) Magnetic - non-magnetic classification accuracy on the four datasets with different thresholds from 0.0 to 1.0 $\mu_B$; (b)-(e) Confusion matrix of the classification on MP, Fe-B, Fe-Co, and Fe-Ti datasets respectively at the threshold 0.5 $\mu_B$.}
  \label{fig:mag_class}
\end{figure}

\clearpage
\begin{figure}[H]
  \centering
  \includegraphics[width=0.8\textwidth]{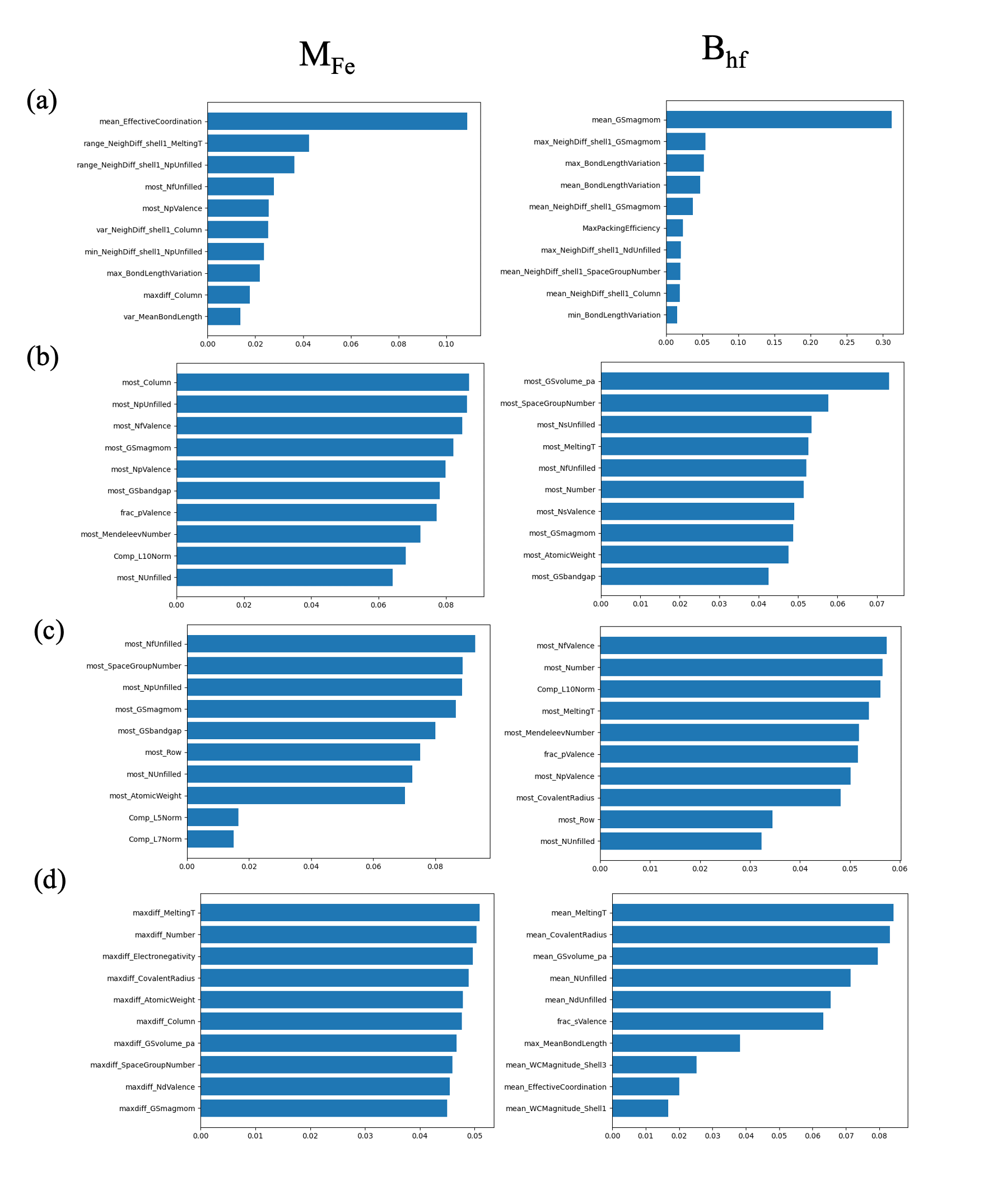}
  \caption{Feature importance analysis for Magpie on $M_{Fe}$ and $B_{hf}$ respectively. (a) - (d) stand for MP, Fe-B, Fe-Co, and Fe-Ti datasets.}
  \label{fig:feature_import}
\end{figure}
\clearpage

\begin{figure}[H]
  \centering
  \includegraphics[width=0.8\textwidth]{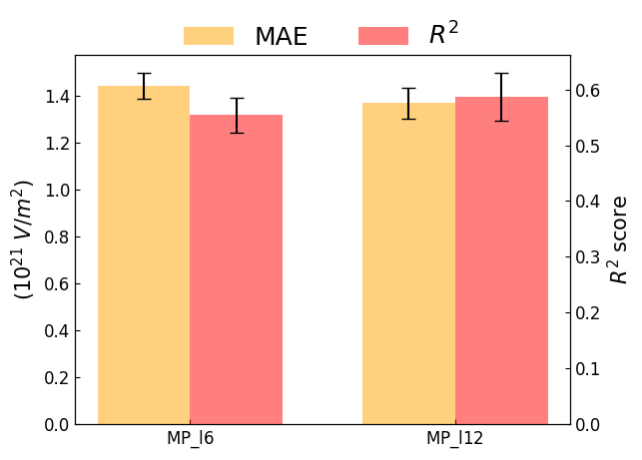}
  \caption{Comparison of EFG regression on MP datasets using different l values in SOAP descriptors, i.e. l=6 and l =12.}
  \label{fig:MP_l6_l12}
\end{figure}


\end{document}